\documentstyle[12pt,aasms4]{article}
\def\propsim{\mathrel{\hbox{\rlap{\hbox{\lower4pt\hbox{$\sim$}}}\hbox{\raise1pt\hbox{$\propto$}}}}}
\begin{document}
\title	{ANALYSIS OF THE SYNCHROTRON EMISSION FROM THE M87 JET}
\author {Sebastian Heinz\altaffilmark{1} \and Mitchell
C. Begelman\altaffilmark{2}}
\affil{JILA, University of Colorado and National Institute for Standards
and Technology, Boulder, Colorado
80309-0440\altaffilmark{3}}
\altaffiltext{1}{email address: heinzs@bogart.Colorado.edu}
\altaffiltext{2}{email address: mitch@jila.Colorado.edu}
\altaffiltext{3}{ also at Department of Astrophysical and Planetary
Sciences, University of Colorado, Boulder}

\begin{abstract}
We propose that the intensity changes and spectral evolution along
the M87 jet can be explained by adiabatic changes to the particle momentum
distribution function and the magnetic field. This is supported by the
lack of any significant variation in the  radio--to--optical
spectral index along the jet and the moderate changes in radio
brightness. Assuming a simple scaling law between magnetic field and
density, we use the deprojection of a 2 cm VLA intensity map by Sparks,
Biretta, \& Macchetto (1996) to predict the spectral evolution along the
jet.

We derive limits for the magnetic field and the total pressure by comparing
our results with the spatially resolved fit to spectral data by Neumann,
Meisenheimer, \& R\"{o}ser (1997) of a model spectrum that cuts off at
$\approx 10^{15} \rm Hz$. To explain the weakness of synchrotron cooling
along the jet , the magnetic field strength must lie below the
equipartition value. Although the inferred pressure in the limit of
nonrelativistic bulk flow lies far above the estimated pressure of the
interstellar matter in the center of M87, bulk Lorentz factors
$\Gamma_{\rm jet}$ in the range of $3 - 5$ and inclination angles
$\theta_{\rm LOS} \lesssim 25^{\circ}$ lead to pressure estimates close to
the ISM pressure. The average best fit magnetic fields we derive fall
in the range of $20 - 40\ \mu\rm G$, departing from equipartition by a
factor $\approx 1.5 - 5$.

This model is consistent with the proposal by Bicknell \& Begelman (1996)
that the knots in the M87 jet are weak, oblique shocks. First--order Fermi
acceleration will then have a minimal effect on the slope of the
radio--to--optical spectrum while possibly accounting for the X-ray
spectrum. 
\end{abstract}

\section{INTRODUCTION}

The jet in M87 has been observed in many wavelength bands, yet  some
puzzles remain about the nature of its emissivity. Polarization  observations in both the optical and radio have shown that
the emission from the jet at wavelengths longward of $\approx 100\ {\rm
\AA}$ is most likely of synchrotron origin (\cite{baade}, \cite{owen}).
Collimated structure can be traced back to $0.01 \rm pc$ from the core
using VLBI (\cite{junor}), and terminates $25''$ away
from the core (2 kpc at the assumed distance of 17 Mpc) in the western
radio lobe, with optical emission  still detectable at this distance. The
jet looks very similar at optical and radio wavelengths, although
Sparks, Biretta, \& Macchetto (1996, hereafter \cite{sparks}) showed that
differences (e.g., in the transverse brightness profiles) do exist.

Far from having a smooth appearance, the jet exhibits a series of  bright
knots at intervals roughly $2''.5$ apart. The
nature of the knots is uncertain, but they are usually attributed to
internal shocks from either flow instabilities (Bicknell \& Begelman 1996,
hereafter \cite{bb}) or variable outflow at the source (\cite{rees}). Both the knots and the interknot regions exhibit power law spectra of
index $\alpha _{R} \approx 0.5$ in the radio, connecting to the optical
data with a power law of index $\alpha _{\rm RO} \sim 0.65$, and steepening
to $\alpha_O$ between $1.2$ and $1.8$ in the optical. This steepening trend
is also found in observations at infrared (e.g., \cite{stocke}) and
ultraviolet (\cite{perola}) wavelengths, as confirmed recently by {\em HST}
observations (\cite{boksenberg}, \cite{sparks}). X-ray observations made
with {\it Einstein Observatory} (\cite{bir:xrayspec}) and {\em ROSAT}
(\cite{meis:rosat} and \cite{Reynolds:rosat}) reveal X-ray emission from 
several spots along the jet (mainly the core and knot A, possibly also from
knots D and B). However, the origin of the X-ray emission is unknown and it
is not clear whether the spectrum breaks between optical and X-ray
wavelengths to a spectral index of $\alpha _{\rm OX} \approx 1.4$, with the
X-ray emission still being of synchrotron origin, or whether the X-ray
emission is produced by a different mechanism, e.g., inverse Compton
scattering or bremsstrahlung.

The onesidedness of the M87 jet has been interpreted as the result of
relativistic beaming of the emission from an intrinsically bipolar jet,
which implies bulk velocities corresponding to Lorentz factors of
$\Gamma_{\rm jet} \gtrsim 2$ with line of sight inclinations of
$\theta _{\rm LOS} \lesssim 30^{\circ}$. This would explain the
existence of the radio lobes on both sides of the core. New infrared
detections of counterjet emission (Stiavelli, Peletier, \& Carollo 1997)
seem to confirm this hypothesis. The Doppler beaming interpretation is
bolstered by proper motion measurements of the knots (\cite{bir:motions}),
which show characteristic velocities $v_{\rm proper} \approx 0.5\ c$ in
knots A and B, with some features exhibiting much larger proper velocities
(a subfeature of knot D appears to show superluminal motion). These
motions broadly support the interpretation of the knots as relatively weak,
{\em oblique} shocks moving down the jet with pattern speeds significantly
smaller than the bulk speed of the flow (\cite{bb}).

One might hope to detect the effects of synchrotron cooling and
relativistic particle acceleration by studying the spectrum as a function
of position along the jet. Such measurements (\cite{sparks}) show
that the radio--to--optical spectral index, $\alpha _{\rm RO}$, is very
nearly uniform, while the optical spectral index, $\alpha _{\rm O}$, is
anti--correlated with the brightness, i.e., the optical spectrum is flatter
in regions of higher intensity. Modeling the optical steepening as a
high-energy cutoff imposed on a power law spectrum, Neumann, 
Meisenheimer, \& R\"oser (1997, hereafter \cite{meis96}) and
\cite{meis:synchspec} find a corresponding correlation between brightness
and cutoff frequency, i.e., a higher frequency at higher intensities. Both
these results are striking in the lack of a strong secular decline in the
cutoff frequency with distance from the core, as would be expected naively
if the steepening were due to synchrotron cooling.

Indeed, these observational results do not compare well with simple
quantitative models of synchrotron cooling in the M87 jet. The usual
assumption of equipartition leads to estimates of the magnetic field of
order $300\ \mu \rm G$, with values up to $500\ \mu \rm G$ in knot A.
The synchrotron lifetime for electrons with Lorentz factors of $\gamma
\approx 10^{6}$, needed to produce the optical emission in a $300\ \mu 
\rm G$ field, is only $2.3\times 10^{10}\ {\rm sec}$, which, for mildly
relativistic bulk velocities (e.g., $0.5\ c$) implies a travel distance of
less than $120 {\rm\ pc}$. But the projected length of the jet is about
$2 {\rm\ kpc}$, and even the distances between 
the most prominent knots are longer than the estimated cooling length. Yet
the spectrum between radio and optical bands remains remarkably constant
along the jet, with only minor variations in the optical spectral index.

This presents a paradox: After a travel distance of $2\ {\rm kpc}$ one
would expect the cutoff to have moved to below $10^{12}\ \rm Hz$, a
factor of $10^{3}$ smaller than what is actually observed. The
discrepancy becomes worse if the magnetic field is stronger than the
equipartition value, as suggested by Owen, Hardee, \& Cornwell (1989) in
order to explain the confinement of the overpressured jet via the magnetic
tension force. (An alternate explanation for the confinement --- that the
radio cocoon surrounding the jet is overpressured with respect to its
surroundings --- has been proposed by \cite{bb}.) Three explanations for
the discrepancy have appeared in the literature:

\begin{enumerate}
\item{First-order Fermi re--acceleration in the knots, interpreted as shocks,
could produce high energy electrons (and possibly positrons) from a
synchrotron-cooled distribution, with a power law of roughly the index
observed. It could also explain the observed X-ray emission. However, since
the power-law index produced by Fermi acceleration is a strong function of
the compression ratio one would need some fine tuning to explain the
observed constancy of the radio--to--optical spectral index, which does not
equal the limiting value for a strong adiabatic shock.  Furthermore, there
does not seem to be a significant amount of cooling between the knots even
at the highest optical frequencies observed, which would be expected for
the assumed $B$-fields and interknot distances.} 

\item{Particles could be transported in a loss-free channel in the interior
of the jet, with the bulk of the emission produced in a thin outer layer of
high magnetic field strength (\cite{owen}). In this picture, the knots and
filaments would be interpreted as instabilities with greatly increased
magnetic field strength wrapped around the jet. The emission would then be
fed by particles from this channel. The radio brightness profiles across
the jet seem to suggest a limb-brightened emission, but a reinvestigation
of the HST observations (\cite{sparks}) shows that the optical emission is
more concentrated to the inner regions of the jet. Also, \cite{sparks}'s
deconvolution places the brightest spots in the jet interior. This argues
against a field-free zone in the jet interior.}

\item{On--the--spot reacceleration by a yet unknown process could
maintain the cut-off particle momentum at the observed level, as has been
proposed by Meisenheimer et al.\ (1996) in a model similar to ours
(see \S \ref{sec:meismod}). This has the advantage of explaining all the
observed features, but invokes unknown physics to explain the apparent lack
of cooling.}
\end{enumerate}

Inspired by the observed correlation between the emissivity variations and
the cut-off frequency, by the newly deconvolved volume emissivity
(\cite{sparks}), and by the new evidence for relativistic bulk velocities 
(\cite{stiavelli}), we propose a simple way of explaining the
observations. The only standard assumption we give up is the assumption of
equipartition, which does not seem to have a very firm physical
foundation anyway. Magnetic fields {\it smaller} than equipartition by
a factor of $1.5 - 3$, coupled with bulk Lorentz factors in excess of $2 -
3$, can readily explain the lack of evidence for synchrotron cooling. In
our model the fluctuations of the cutoff frequency are produced by weak
shocks, so that the influence of the compressions on the plasma
distribution function can be considered to be adiabatic. As a result, we
are able to explain the general behavior of the cutoff reasonably well;
{the magnetic fields we derive from our own fits to the data are below
equipartition with a convincing level of confidence, but the inferred total
pressures do not necessarily need to exceed the equipartition
values}. Additionally, for relativistic jets the equipartition values for 
$B$-field and pressure are less than in the nonrelativistic limit, so the pressures we derive can fall below the equipartition value in the
nonrelativistic case.

We organize the paper as follows.  In \S ~\ref{sec-adiabatic} we present a
nonrelativistic treatment of the synchrotron emissivity, taking into
account cooling but assuming that the particles and fields respond
adiabatically to changes in the flow density. Once relativistic effects are
incorporated into the treatment in \S ~\ref{sec-relativistic}, we use the
data of \cite{sparks} and \cite{meis96} to constrain the magnetic field
strength of the jet. Section \ref{sec-discussion} discusses confinement and
stability of the jet in the light of the pressures derived from the results
of \S ~\ref{sec-relativistic}, the production of X-ray emission in knot A,
polarization, and limits on the particle acceleration site; and \S
~\ref{sec-conclusion} gives a brief summary of the results and future
prospects.

\section{ADIABATIC EFFECTS ON SYNCHROTRON EMISSION}
\label{sec-adiabatic}

Our model rests on the hypothesis that Fermi acceleration is unnecessary to
explain the fluctuations of radio--to--optical emissivity and cutoff
frequency along the M87 jet.  Given certain assumptions about the
orientation and degree of disorder in the magnetic field, and the degree of
anisotropy permitted in the relativistic electron distribution, we can
relate changes in both the emissivity and the cutoff frequency uniquely to
changes in the density of the jet fluid.  These adiabatic effects are
readily combined with the effects of synchrotron cooling
(\cite{coleman}). In effect, given the emissivity map of \cite{sparks}, we
can {\it predict} the run of the cutoff frequency along the jet, and
vice-versa.  Since we also have NMR's observations of the cutoff
frequency as a function of position, our adiabatic model is subject to a
powerful self-consistency check.

Observationally, the main changes in emissivity and cutoff frequency are
rather localized, and associated with the positions of the knots. These
small-scale fluctuations therefore provide the strongest check on our
assumption of adiabaticity. The {\em large-scale} trends then determine the
best fit to the magnetic field strength. We have already seen that the
apparent lack of a large-scale synchrotron cooling trend is incompatible
with a magnetic field strength as large as the mean equipartition value
(see Fig. \ref{fig:freqfit}). Neglecting relativistic and projection
effects, it would require a magnetic field as low as $ 25\ \mu \rm G$ to
obtain a cooling length of $2\ {\rm kpc}$ at optical frequencies. This is
an order of magnitude smaller than the mean equipartition field and would
require a total (particle + magnetic) pressure of $1.2 \times 10^{-7}\
{\rm dyn}\ {\rm cm^{-2}}$ to produce the observed average amount of
synchrotron emission, compared to an equipartition value of $\approx 4.0
\times 10^{-9}\ {\rm dyn}\ {\rm cm^{-2}}$.

As noted earlier, the physical basis of equipartition is
weak. Estimates of $B$-fields in radio hot spots based on the synchrotron
cooling time indicate that in some cases equipartition might be correct up
to a factor of $\sim 2$ (\cite{meis:equi}), but the conditions in the jet
might very well be different from those in the lobes. We are therefore
free to consider the magnetic field strength to be a free
parameter. Applying our adiabatic model to the observational data, we can
derive an estimate for the magnetic field strength. In the case of the M87
jet, this estimate lies {\em below} equipartition, even when relativistic
effects are taken into account (\S ~\ref{sec-relativistic}).

Our model for the evolution of the particle distribution function follows
that of Coleman \& Bicknell (1988). We assume that pitch angle
scattering due to plasma micro--instabilities keeps the particles close to an
isotropic distribution in the fluid rest frame. In the absence of cooling,
this would imply that the relativistic electrons respond to compressions
like a $\gamma_{\rm adiabatic} = 4/3$ (i.e., ultrarelativistic) fluid, but
this behavior will be modified by synchrotron cooling. Because the magnetic
field is frozen into the plasma, its strength should change as the plasma
density fluctuates along the jet. 

Depending on the orientation and the degree of disorder of the field, its
variation will depend roughly on the density change to some power $\zeta
$: $B(r)\propto \varrho(r)^\zeta $, where $\varrho$ is the proper particle
density and $r$ is the distance from the core. For a completely
disordered magnetic field, $\zeta =\frac{2}{3}$, whereas for a homogeneous
field the power depends on the orientation of the field with respect to the
compression normal, with $\zeta =1$ for an orthogonal orientation, $\zeta
=0$ for parallel orientation. Since the polarization of the jet is of order
10\%-20\% in the interknot regions, compared to the maximum polarization of
70\% for a homogeneous field, it is likely that the magnetic field has
a disordered component, so it is reasonable to assume $\zeta $ to be of order
$\frac{2}{3}$ (but see \S \ref{sec:polar}). The ordered $B$-component is
aligned with the jet axis almost everywhere except in the brightest knots,
as can be seen from polarization measurements (\cite{bir:book}). We do not
know its orientation with respect to the compression normal, because the
orientations of the (presumably oblique: see \cite{bb}) shocks are
unknown. However, our results are not very sensitive to what the actual
value of $\zeta $ is, as we will show later. We therefore make the
simplifying assumption of a single exponent describing the field
variations. Using the scaling relation for the synchrotron emissivity of a
power law momentum distribution ($f(p)\ {\rm d}^3p \propto p^{-a}\ {\rm
d}^3p \propto p^{-a+2}\ {\rm d}p$, corresponding to a spectral index of 
$\alpha = \frac{a - 3}{2}$) under adiabatic compression (e.g.,
\cite{coleman}):
\begin{equation} 
j\propto (B\sin\vartheta)^{1+\alpha}\varrho ^\frac {2\alpha +3}{3}\nu ^{-\alpha},
\label{eq:emissivity}
\end{equation}
where $\vartheta$ is the angle between magnetic field and line of sight,
we can express the field relative to its value at $r_{\rm 0}=0''.5$, the
(arbitrary) injection point at which we start the calculation, as a
function of the emissivity ratio $j/j(r_{\rm 0})$:
\begin{equation}
B=B(r_{\rm 0})(j/j(r_{\rm 0}))^{\xi}
\label{eq:magnetic} 
\end{equation}
where $\xi\equiv\left[1+\alpha+(2\alpha+3)/(3\zeta )\right]^{-1}$. A
necessary condition for this approach to be valid (in addition to the
assumed isotropy and the absence of Fermi acceleration) is the assumed
steady-state injection of relativistic particles and fields by the central
engine, which allows us to relate densities and fields at each $r$ to the
corresponding values at the injection point $r_0$ at a given instant of
time.

Is it plausible to neglect Fermi acceleration in the shocks that comprise
the knots? Except for knot A, the brightness changes along the jet are
moderate. In knot A, the brightest feature, the emissivity changes by a
factor of order $10$ (\cite{sparks}), which, if entirely due to a sudden
compression of the plasma, can be produced by a proper compression {ratio of
$r\approx 2.7$} as measured in the respective rest frames of the plasma if
we take $\zeta=\frac{2}{3}$. For the other knots we infer smaller density
contrasts, $r \approx 1.5$. Consistent with this observation, we will
henceforth take the knots to be weak shocks, in accordance with the
suggestion by \cite{bb} that the knots are highly oblique (and therefore
weak) shocks. Thus, because Fermi acceleration leaves the spectral index
unchanged if the shock is weak enough, we will henceforth neglect its
effect on the cutoff frequency. Section \ref{sec-discussion} discusses
Fermi acceleration in more detail, with particular focus on the possibility of
Fermi acceleration occurring in knot A.

Furthermore, because the shocks are believed to be oblique, we take the
fluid velocity to be constant to first order, both in magnitude and
direction. For the shock jump conditions in the non--relativistic limit
(which we consider in this section) the velocity component perpendicular to
the shock plane $v _{\perp}$ is inversely proportional to the density, thus
for a proper compression ratio of $2.7$, as seen in knot A, the
perpendicular velocity component should change by a factor of $0.37$. For
highly oblique shocks, $v _{\perp}$ is small compared to $v _{\parallel}$
and the velocity will not change significantly. Moderate changes in
velocity would be easy to incorporate in principle; yet with our current
ignorance of the velocity field and shock parameters, such a level of
detail is unwarranted. We will comment on the validity of this assumption
in \S ~\ref{sec-relativistic}. We also postpone a treatment of the motion
of the knots until \S ~\ref{sec-relativistic}, and assume them to be
stationary for the rest of this section.

With these assumptions we are ready to calculate the downstream cutoff
frequency for a given initial cutoff momentum in the injected particle
distribution and a given $B$-field at $r_{\rm 0}$. We use the transport
equation as presented by Coleman \& Bicknell (1988):
\begin{equation}
\frac{d f}{d t}+\frac{1}{3\varrho}\frac{d\varrho}{dt}p\frac{\partial
f}{\partial p}=Ap^{-2}\frac{\partial}{\partial p}(p^{4}f)
\label{eq:transport}
\end{equation}
written in the rest frame of the fluid. Here, $f=f(p)$ is the electron
distribution function and $A=\frac{4e^4}{9m_{\rm e}^{4}c^{6}}B^{2}$ is the
synchrotron loss term. The equation is valid for the assumed case of
isotropy and negligible Compton losses (for a brief discussion of Compton
losses see \S ~\ref{sec-discussion-confinement}). The solution of  this
equation is
\begin{equation}
f(p(r))=f _{\rm 0}(p _{\rm 0}) \left(\frac{p _{\rm 0}\ \varrho^{\frac{1}{3}}}{p\
\varrho_{\rm 0}^{\frac{1}{3}}}\right)^{4}
\label{eq:transsolution}
\end{equation}
where $f _{\rm 0}$ is the injected momentum distribution and
\begin{equation}
p(r)=\frac{(\varrho /\varrho
_{\rm 0})^{\frac{1}{3}}p_{\rm 0}}{(1+p_{\rm 0}\int_{t(r_{\rm 0})}^{t(r)}A(t')(\varrho (t')/\varrho
_{\rm 0})^{\frac{1}{3}}dt')} 
\label{eq:solution}
\end{equation}
(see \cite{coleman}) where the subscript 0 denotes the values at
injection point $r_{\rm 0}$. 

Equation (\ref{eq:solution}) describes how the momentum of a given particle
changes along a streamline. Thus, if the distribution initially cuts off at
$p _{\rm c,0}$, we can calculate the cutoff momentum $p _{\rm c}(r)$
downstream. Because in our model the density $\varrho (r)$ is proportional
to $B(r)^{\frac{1}{\zeta}}$ and because we know the scaling of $B$ with $r$
from equation (\ref{eq:magnetic}), we can eliminate $\varrho$ and $B/B
_{\rm 0}$ from equation (\ref{eq:solution}). The remaining parameters are $B
_{\rm 0}$, $p _{\rm c,0}$, and ${j(r)}/{j _{\rm 0}}$, the latter being provided by the
Sparks et al.\ data.

The cutoff momentum $p _{\rm c}$ is related to the observed cutoff frequency
$\nu _{\rm c}$ by the expression
\begin{equation}
\nu _{\rm c}=\frac {3e}{4\pi m_{\rm e}^{3}c^{3}}p _{\rm c}^{2}B\
\sin{\vartheta} .
\label{eq:critical}
\end{equation}
Equation (\ref{eq:critical}) contains another parameter, $\vartheta $,
the angle between the line of sight and the magnetic field. For now we
shall set the factor $\sin{\vartheta}\approx 1$, which is valid for
disordered fields, since 1) the regions in which the field is perpendicular
to the line of sight have the highest emissivity, and 2) assuming randomly
oriented fields,  half of the field orientations lie in the range from
$60^{\circ}$ to $90^{\circ}$ to the line of sight,
i.e., $\sin{\vartheta}\geq 0.866$. Thus the cutoff frequency is mainly
determined by field orientations close to $90^{\circ}$\ or
$\sin{\vartheta} \approx 1$.

We can now determine the free parameters $B _{\rm 0}$ and $p _{\rm c,0}$ by
applying a least chi-squared method to fit the observed cutoff frequency
$\nu _{\rm c, {\rm obs}}$ with the value determined from equations
(\ref{eq:solution}) and (\ref{eq:critical}), using the emissivity map
$j(r)/j _{\rm 0}$ provided by \cite{sparks}. We prefer to average the
emissivity across the jet, which minimizes small scale variations probably
due to the deprojection procedure. Because \cite{meis96} also averaged
across the jet, this seems to be the most appropriate way of calculating
the cutoff frequency. Figure \ref{fig:freqfit} (calculated for a bulk
Lorentz factor of $\Gamma_{\rm jet} =1.1$, a radio spectral index of
$\alpha _{\rm R} =0.5$, $\theta _{\rm LOS}=90^{\circ }$, and $\zeta =\frac{2}{3}$)
shows the observed cutoff frequency $\nu _{\rm c, {\rm obs}}$ (vertical
bars) with error bars and the best fit curve (solid line), which seems to
reproduce the scaling of the cutoff frequency reasonably well. {(The radio
spectral index seems to break to $\alpha \sim 0.65$ at $\sim 10$ GHz, so we
have used both $\alpha=0.5$ and $0.65$ in our fits with insignificant
differences in the average parameters but smaller chi-squared for $\alpha
\sim 0.65$; see \S \ref{sec-relativistic}.)} For comparison the plot also
shows the best fit cutoff frequency for equipartition $B$-fields (dashed
line). The mean $B$-field is of order $10\ \mu\rm G$, even smaller than the
zeroth order estimate made at the beginning of this section. Assuming
(arbitrarily) a lower cutoff at $\nu =10^{7}\ {\rm Hz}$ and the observed
high-frequency cutoff at $\nu\approx 10^{15}\ {\rm Hz}$ yields an average
total pressure of ~$8\times 10^{-8}\ {\rm dyn}\ {\rm cm^{-2}}$ for the
given parameters, compared to an equipartition value of $p\approx 3\times
10^{-9}\ {\rm dyn}\ {\rm cm^{-2}}$. In calculating absolute values for both
pressure and $B$-field, projection (i.e., foreshortening and length scale)
effects must be taken into account, since the emissivity was derived for a
side-on view of the jet. This introduces a factor of $\sin{\theta _{\rm
LOS}}$ in intrinsic emissivity and pressure for a given magnetic field and
a factor of $(\sin{\theta _{\rm LOS}})^{\frac {4}{7}}$ in equipartition
pressure.

It seems that the proposed modest compressions can account for the
fluctuations seen in the spectral cutoff, and the large scale decrease in
$\nu _{\rm c}$ is well reproduced. It would be helpful to determine both the
emissivity and the spectral index maps from the same method and data, thus
eliminating errors due to different reduction procedures. We comment on the
deviations and uncertainties in this fit in \S ~\ref{sec-relativistic}.
It is important to note that this technique should be independent of what
the actual shape of the particle distribution is, because it simply tracks
the behavior of a single feature in the spectrum, which could be identified
with either a break or a cutoff.

As seen in this section, one runs into problems with the jet pressure for
nonrelativistic bulk velocities. Also, the observed mildly relativistic
proper motion of the knots and the jet's onesidedness favor a relativistic
interpretation, as do the knot spacing and morphology (\cite{bb}). The results
indicate that Lorentz factors of order $2 - 5$ fit the observations
best. In the next section we will investigate the effects of these
suggested relativistic bulk velocities.

\section{RELATIVISTIC EFFECTS} 
\label{sec-relativistic}

Relativistic motions not only explain the onesidedness of the M87
jet, but also help to solve the synchrotron cooling problem mentioned in the
introduction. The travel time in the electron rest frame is reduced by a
factor $\Gamma_{\rm jet}$ due to time dilation, the intrinsic emissivity
is reduced by a factor $D^{2+\alpha}$ (where $D$ is the Doppler factor
$D=\left[\Gamma_{\rm jet} (1-\beta\cdot \cos{\theta _{\rm
LOS}})\right]^{-1}$), and the intrinsic cutoff frequency is Doppler
shifted downward by a factor $D$. As a result, the apparent synchrotron
lifetime can be a significant underestimate of the intrinsic value.

Biretta (1993) estimates the lower limit on the jet--to--counterjet
radio brightness ratio to be $\geq 150 - 380$ (the higher value corresponds
to the assumption that jet and counterjet have identical appearance).
Based on this limit we adopt line--of--sight angles $\leq 35^{\circ}$ and
Lorentz factors $\geq 2$. Note that even Doppler factors smaller than unity
can lead to a large jet--to--counterjet brightness ratio, as the counterjet
brightness is severely reduced for large $\Gamma_{\rm jet}$.

We repeat the analysis of \S ~\ref{sec-adiabatic} using the emissivity
profile of \cite{sparks}, this time corrected for Doppler boosting and
projection effects. Again, for fitting the cutoff frequency only emissivity
ratios are important, so these corrections do not change the fitting
procedure as long as changes in the bulk velocity can be neglected. We are
confident that at least the direction of the flow is not changed
significantly before knot C. The only knot for which such effects could be
important is knot A, because it displays a jump in emissivity of $11$,
whereas in the other knots the emissivity is increased by a factor of order
$3$ only, which implies very moderate compressions. Using the relativistic
continuity equation for an oblique shock we have estimated
the post--knot A Lorentz factor to be $\Gamma_{jet,A+} \sim 3$ for a
pre--knot A $\Gamma_{jet,A-}=5$, a compression ratio of 3 and intrinsic
obliquities $\sim 60^{\circ}$. Although the change in $\Gamma_{jet}$ might
seem large at first glance, the impact such a velocity change has on our
fits is not large, as we will explain below. We modify equation
(\ref{eq:transport}) to follow the electron distribution in the {\em fluid
rest frame} by replacing $t$ with $\tau$ and $\frac {d}{d t}$ with
$\frac{d}{d \tau}$, where $\tau $ is the proper time. The same changes
apply to equation (\ref{eq:solution}). Strictly speaking, the fluid frame
is not an inertial frame and we would have to include accelerational terms
into the equation, introducing an anisotropy. But our assumption should be
adequate, provided that isotropization takes place over short enough
scales.

Treating the response of the distribution function to compressions as
adiabatic and assuming isotropy (i.e., an adiabatic index of $\frac
{4}{3}$) we calculate the changes in $B$ and cutoff momentum (measured
in the fluid frame) from the emissivity changes in the fluid frame. The
substitution $t\rightarrow \tau$ takes care of the time dilation
effects.

In order to incorporate the observed motion of the knots, we must correct
$A(\tau ')$ and $\varrho (\tau ')$ in equation (\ref{eq:solution}) for light
travel time effects between the source and the observer. This is because
the knots move during the time it takes a particle to travel from 
$r _{\rm 0}$  to $r$.  We need to know the ratios $B/B_{\rm 0}$ and
$\varrho/\varrho_{\rm 0}$ experienced by a particle as a function of {\em
proper time} $\tau '$ in order to be able to do the integration in
equation\ (\ref{eq:solution}). Because we infer the relative values of $B$
and $\varrho$ at a given position from the emissivity ratio $j(r _{\rm
fluid})/j _{\rm 0}$, it is important to know the velocity of the emissivity
pattern.

We assume that the pattern of density and field fluctuations retains its
shape and moves along the jet at a fixed speed $v_{\rm pattern}$, taken
to be smaller than $v_{\rm fluid}$ and set equal to $0.55\ c$
everywhere in our calculations for simplicity. If the present field
distribution (i.e., at a given time $t=0$ in our frame, corrected for
light travel time effects) is expressed by $B(r)$, then the field at
time $-t'$ and position $r'$ is given by $B(r' + v_{\rm pattern} t')$.
Now, for a particle currently at $r$, the equation of motion is $r' = r -
v_{\rm fluid} t'$. Therefore, the field distribution experienced by the
particle as a function of time is $B[r - (v_{\rm fluid} -v_{\rm pattern})
t']$. Appropriate modifications to equation  (\ref{eq:solution}) are
straightforward. The effect of the pattern speed on the result is not very
dramatic, reducing $\chi^{2}_{\rm min}$ by about $6\%$.

With this set of assumptions we can once again proceed to integrate the
modified equation (\ref{eq:solution}) for various $\theta _{\rm LOS}$ and
$\Gamma _{\rm jet}$. Using a minimum chi-square routine we can determine
the best--fit values for $B_{\rm 0}$, and $p_{\rm 0}$. Relation
(\ref{eq:magnetic}) then yields $B(r)$.

Figure \ref{fig:contourplot} shows a chi-square plot for $\Gamma_{\rm
jet}=3$, $\theta _{\rm LOS}=25^{\circ}$, and $\zeta =\frac{2}{3}$. The
equipartition value for the average $B$-field is shown as a shaded area
at $89\ \mu \rm G$. The upper limit on $B_{\rm mean}$, set by $2\chi
^{2}_{\rm min}$ contours, lies at $49\ \mu{\rm G}$, 75\% above the best
fit value, $B_{\rm mean}=28\ \mu{\rm G}$. The lower limit set by
$2\chi^{2}_{\rm min}$ is $\approx 5\ \mu{\rm G}$, 80\% below the best fit
value. The average equipartition field of $B_{\rm mean}\approx 89\ \mu\rm
G$ lies above even the $5 \chi^{2}_{\rm min}$ contour. The lower limit on
$B$ is not nearly as strict, due to the fact that cooling is not dominant,
i.e., we can produce a similar spectral behavior by reducing the magnetic
field and increasing the particle energy, which produces the
tear-shaped appeareance of the contours. Thus, strictly speaking, the best
fit values for the $B$-field should be regarded as upper limits.

The reduced $\chi^{2}_{\rm min}$ values (i.e., $\chi^{2}_{\rm min}$ divided
by the number of degrees of freedom) fall above 44, which is uncomfortably
high. However, because we do not have formal errors for the emissivity
deprojection by \cite{sparks}, which will introduce a significant
uncertainty, a high value for $\chi^{2}_{\rm min}$ is not all that
discouraging. Estimating the average uncertainty in the emissivity by
comparing the averaged emissivity to that derived from taking only a slice
along the jet yields an uncertainty of order 50\%, which leads to
uncertainties in the predicted cutoff frequency of roughly 20\%. This
is significantly higher than the formal error in \cite{meis96}'s data and
will reduce the $\chi^{2}_{\rm min}$ by a factor of approximately 10.

The $\chi^{2}_{\rm min}$ values are dominated by the region beyond knot
A. The post--knot A residuals in our fit are not larger than the
residuals in the pre--knot--A region, but because the post knot A region is
brighter, the error bars on the measured cutoff frequency are smaller,
which increases the $\chi^{2}_{min}$. The deprojection procedure, which
assumed an axially symmetric flow, breaks down beyond knot A, which will
introduce significant uncertainty. Also, non--uniformities in the
emissivity could lead to large errors if the optical emission peaks at
different locations than does the radio emission. Field orientation effects
and changes in $\Gamma_{\rm jet}$ and $\theta_{\rm LOS}$ might also
contribute to the error. We performed the same procedure just out to knot A
and found that, with the {\em same} parameters, the reduced
$\chi^{2}_{min}$ shrinks to 13. Leaving $B$  and $p_{c}$ as {\em free}
parameters reduces $\chi^{2}_{min}$  to 10, but also reduces the $B$-field
significantly. Because in this case the algorithm mainly fits the region
around knot A (where the error bars are smallest), we cannot expect the
global run of $\nu_{c}$ to have significant impact on the fit, which would
be necessary to extract information about the average magnetic field. We
conclude that the reproduction of fine detail is not satisfactory in the
region beyond knot A. However, the gross run of $\nu_{c}$, which is
principally responsible for constraining our parameters, is reasonably well
reproduced.

The best--fit average magnetic field $B_{\rm mean}$, as a function of
$\Gamma_{\rm jet}$ and $\theta_{\rm LOS}$, is plotted in Figure
\ref{fig:Bmean}. Figure \ref{fig:Brels1p5} shows the average ratio
$\langle {B} / {B_{\rm eq}}\rangle $ as a function of $\Gamma _{\rm jet}$
for $\theta _{\rm LOS}=15^{\circ}$ to $30^{\circ}$ in increments of
$5^{\circ}$, and the area corresponding to the limiting jet--to--counterjet
brightness ratio of 150 - 380. Note that the equipartition magnetic field
has to be corrected by a factor of $D^{-\frac {2+\alpha }{3+\alpha
}}(\sin{\theta _{\rm LOS}})^{\frac {1}{3+\alpha}}$ for projection and
Doppler boosting of the emissivity; this has already been taken into
account in the  figure. Clearly, for $\Gamma _{\rm jet}$ in the range $3-5$
and $\theta _{\rm LOS}\lesssim 25^{\circ}$ the departure from equipartition
is not very large (roughly a factor of $0.2 < \langle {B} / {B_{\rm
eq}}\rangle < 0.6$).

In order to test the dependence of the best fit $B_{\rm 0}$ on the
parameter $\zeta$ we have calculated the same curves for $\zeta =1$ and
$\zeta =\frac{1}{15}$. Figure ~\ref{fig:Berr} shows the fractional
deviation $\left(\frac{\Delta B}{B}\right)_{\zeta}\equiv\frac{B_{\rm
\zeta_{1}}-B_{\rm \zeta_{\rm 2}}}{B_{\rm \zeta_{\rm 1}}}$ from the
$\zeta_{1} =\frac {2}{3}$ curve for models with $15^{\circ} \leq
\theta_{LOS} \leq 30^{\circ}$. The deviation is small compared to the
expected errors introduced by the simplifications we made and to the range
in $B$ allowed by our minimum chi-square procedure, at most $12 \%$ for
$\zeta_{2} =\frac{1}{15}$ and small $\Gamma_{\rm jet}$. This is not a very
reasonable value for $\zeta$ in any case, because the field has a random
component, thus $\zeta$ should be higher, and the probability of the field
being in the shock plane (thus having $\zeta =1$) is twice as high as for
the field being normal to the shock. We conclude that our ignorance of the
precise behavior of $B$ under compression is not a serious obstacle to the
application of our model.
 
We also tested the impact a change in $\Gamma _{jet}$ at knot A might have
on our results. As we mentioned earlier, the best fit $B$-field values we
derive are upper limits. This is the reason why a change in
$\Gamma_{jet}$ at knot A does not change our results significantly:
generally, lower Lorentz factors require lower fields to explain the
observed lack of cooling. If the jet is slowed down beyond knot A, we will
need lower average fields to fit this region. However, lowering
the field does not change the quality of the fit much (the $\chi ^2$ is
essentially unchanged), so the global field strength is simply set by the
region with the lower Lorentz factor, $\Gamma_{jet,A+}$ (the relative
scaling of $B$ is still determined from equation \ref{eq:magnetic}, taking
relativistic beaming into account). We have introduced by hand a change of
$\Gamma_{jet}$ at knot A into our model (we solved the continuity equation
at the shock, assuming an obliquity of $60^{\circ}$, for the velocity
change that would reproduce the observed emissivity jump of 11, including
relativistic beaming and adiabatic compression), and calculated the
fractional deviation
$\left(\frac{\Delta{B}}{B}\right)_{\Gamma}\equiv\frac{|B_{\rm uniform} -
B_{\rm break}|}{B_{\rm uniform}}$ of the derived averaged $B$-field. Here,
$B_{\rm uniform}$ is the best fit average $B$-field derived for uniform
$\Gamma_{jet}$ and $B_{\rm break}$ is the best fit field for a jet slowing
down from $\Gamma_{jet,A-}$ to $\Gamma_{jet,A+}$ at knot A. Figure
\ref{fig:Berr} shows $\left(\frac{\Delta B}{B}\right)_{\Gamma}$ for uniform
jet models with $\Gamma_{jet}$ set to either $\Gamma_{jet,A-}$ or
$\Gamma_{jet,A+}$ (filled light and dark grey regions, respectively). The
latter is always less than 18\% for the parameter range we used. Note that
for post--knot--A $\Gamma_{jet,A+}$'s above 5, the pre--knot--A
$\Gamma_{jet,A-}$ exceeds 8.5, thus $\Gamma_{jet,A+} \gtrsim 7$
can be ruled out on the basis of gross energy balance arguments (see next
section).

For completeness we have shown the deviation of the best fit average
$B$-field $\left(\frac{\Delta B}{B}\right)_{\alpha} \equiv
{\frac{|B_{\alpha_{1}}-B_{\alpha_{2}}|} {B_{\alpha_{1}}}}$ for a 2 cm
radio spectral index of $\alpha_{2} = 0.65$ instead of $\alpha_{1} = 0.5$
as the black region in Figure \ref{fig:Berr}. One can see that the
difference is negligible compared to other uncertainties.

\section{DISCUSSION} 
\label{sec-discussion}

In the preceding sections we demonstrated that a) magnetic fields slightly
below equipartition and b) moderately relativistic effects are able to
explain the general behavior of the spectrum in M87. In this section we
will examine the confinement properties of the jet and compare our model
with a previous model by Meisenheimer et al.\ (1996). We also comment on
the production of X-ray emission in knot A, and on the consistency of our
model with polarization measurements, and we estimate the minimum distance
from the core at which particle acceleration has to occur.

\subsection{\em Confinement}
\label{sec-discussion-confinement}
Naturally the question arises whether the jet can be confined under the
conditions we proposed above. The usual assumption for a jet to be confined
is that it is in pressure equilibrium with its surroundings. Alternatively,
one could imagine the jet to be freely expanding into an underpressured
surrounding medium. 

\cite{bb} argue that in order to produce shocks via
Kelvin-Helmholtz instability, some interaction between jet and surrounding
medium has to take place, as opposed to a free expansion scenario. They
also show that the minimum Lorentz factor $\Gamma_{\rm jet}$ for a freely
expanding jet with no cold matter content is at least $13$, much higher
than the values we have used above. Because we would most certainly fall
out of the beaming cone for such a high $\Gamma_{\rm jet}$ jet, the
intrinsic emissivity would be much higher than the observed value. As
\cite{bb} point out, the energy flux of the jet would far exceed the 
estimates made on the basis of the expanding bubble the M87 jet blows into
the ISM. A $\Gamma_{\rm jet}$ that high would also raise questions about
the location at which the jet is decelerated to nonrelativistic velocities,
and seems inconsistent with the claimed detection of IR counterjet emission
by Stiavelli et al. (1997). We can therefore rule out the picture of a
freely expanding jet. As a consequence we need a mechanism to provide
confinement, i.e., we need to set the jet pressure in relation to the
ambient pressure.

The ambient gas pressure in the center of M87 has been derived by
\cite{white} from fitting cooling flow models to the {\em Einstein} 
X-ray observations. The values they find fall into the range $p_{\rm ISM} =
1\times 10^{-10}\ \rm dyn\ cm^{-2}\ $ to $\ p_{\rm ISM}=4\times 10^{-10}\
\rm dyn\ cm^{-2}$. It is important to note that the pressure of the
interstellar medium in M87 might not be representative of the pressure of
the immediate environment of the jet. In fact, \cite{bb}'s analysis of the
helical Kelvin-Helmholtz instability leads to the conclusion that the
ambient medium of the jet is significantly overpressured with respect to
the interstellar medium in M87. Note also that the pressure in the knots
might well exceed the ambient pressure without losing confinement, as long
as the average pressure does not.

We have calculated the average total pressure $p_{\rm mean}$ in the jet
from the averaged emissivity and the best--fit $B$-field for various
angles and Lorentz factors, as shown in Figure ~\ref{fig:logpressure}. The
assumptions we have made in constructing this plot are analogous to those
of \cite{sparks}, who (arbitrarily) assumed a lower cutoff at $10^{7}\rm
Hz$, a high energy cutoff at $10^{15} \rm Hz$, a spectral index of 
$\alpha _{RO}=0.5$, and equipartition between heavy--particle and electron
energy. (Note: because the spectrum is steeper than $\alpha_{R}=0.5$
above 10 GHz, our estimate of the pressure is likely to be an
overestimate.) We have also assumed isotropic emission in the plasma rest
frame by using an average value of $\vartheta=54^{\circ}$ for the term
$\sin^{1+\alpha}{\vartheta}$ in the emissivity equation
(\ref{eq:emissivity}).

For comparison we have also calculated the equipartition pressure and
plotted the ratio of pressure to equipartition pressure in Figure
\ref{fig:relpressure}. It is obvious that we are far above equipartition
for small values of $\Gamma_{\rm jet}$ and large $\theta_{\rm LOS}$,
but as we approach the favored range of $\Gamma_{\rm jet} \geq 3$ and
$\theta_{\rm LOS} \lesssim 30^{\circ}$, $p_{\rm total}$ approaches the 
equipartition value. The exact value of the pressure depends critically on
the details we put into the model spectrum. For a jet composed entirely of
electrons and positrons, the pressure would go down by a factor of
$\frac{1}{2}$, whereas the equipartition pressure would only decrease by
a factor of $(\frac{1}{2})^{\frac{4}{7}}=0.67$. The lack of information
about the low--frequency spectrum inhibits any statements about the
low--energy particle distribution. However, it is safe to assume that the
power law does not continue down to non-relativistic energies.

It is obvious that a magnetic field far below equipartition alone cannot
explain the behavior of the jet --- it might account for the spectral
changes but it requires the pressure to be much higher than that of the
surrounding medium. Field orientations close to the line of sight will
lead to an underestimate in emissivity, pressure, and intrinsic cutoff
frequency and will require even lower magnetic fields and even higher
pressures in order to prevent significant cooling. This changes as we
increase $\Gamma_{\rm jet}$: the inferred pressures are close to
the value for the interstellar medium in M87 as derived by \cite{white},
for $\Gamma_{\rm jet}\approx 3 - 5$ and $\theta_{\rm LOS}\lesssim
25^{\circ}$. This, combined with the possible overpressure of the jet's
immediate environment relative to the ISM, leads to the conclusion that
there is no confinement problem.

For small values of the magnetic field, one might ask if inverse Compton
losses become dominant. A simple order of magnitude estimate shows that
this is not the case. The ratio of synchrotron to inverse Compton loss
timescales is equal to the ratio of photon energy density to magnetic field
energy density (\cite{rybicki}). The magnetic field strength has been
estimated above. To derive an estimate of the photon energy density
produced by the synchrotron emission, we normalize to the radio luminosity
corrected for beaming and projection effects and integrate over an
$\alpha _{\rm RO}=0.65$ radio--to--optical power law that cuts off at $10^{15}$
Hz. This shows that the inverse Compton lifetime due to just the
synchrotron radiation field of the jet is roughly an order of magnitude
longer than the synchrotron lifetime for the parameter range we suggested
above. The starlight background at the center of M87 also contributes to
the photon energy density. Using an isothermal sphere profile, normalized
to the total luminosity of M87, we arrive at a central photon energy
density roughly an order of magnitude smaller than that of the magnetic
field, small enough to justify the assumption of negligible Compton
losses.

\subsection{\em Comparison to Earlier Models}
\label{sec:meismod}
It is instructive to compare our model to an earlier {\em ad-hoc} model by
Meisenheimer et al. (1996, see also \cite{bir:book}), which bears a lot of
similarity to our model. They start from the same assumption that the
spectral changes along the jet can be explained by simple compressions and
assume that the cutoff {momentum} $\gamma_{c}$ is almost constant along the
jet, parameterizing it as a function only of the transverse jet radius
(measured from the 2 cm radio map): $\gamma_{c}\propto R^{-\frac{1}{3}}$
--- note that for an adiabatic compression of the plasma {\em transverse}
to the flow $\gamma_{c}$ goes as $R^{-\frac{2}{3}}$. {They take the
$B$-field to consist predominantly of a toroidal component, $B_{\Phi}$, and
hold the poloidal component $B_z$ fixed.} They determine the {longitudinal}
compression ratio of $B_{\Phi}$ and of the particle density $n$ from their
fit to the cutoff {frequency} $\nu_{c}$ with equation
\ref{eq:critical}. However, in an adiabatic compression, the cutoff
momentum varies as $n^{\frac{1}{3}}$ and will therefore be affected by
{longitudinal} compressions as well (here is where our assumption of a
disordered field allows us to determine a relation between density and
magnetic field, so we can solve equation \ref{eq:critical} uniquely for
$B$). They neglect the fact that the synchrotron emissivity is enhanced in
adiabatic compressions by $n^{1+{\frac{2}{3}} \alpha}\cdot B^{1+\alpha}$
(equation \ref{eq:emissivity}) rather than $n\cdot B^{1+\alpha}$. Since
the shocks might well be oblique, their assumptions that $B_{\Phi}$ scales
as the longitudinal compression ratio and {that $B_z$ is constant}
might also not be valid.

Meisenheimer et al.\ (1996) favor an intrinsically onesided,
subrelativistic jet, viewed close to perpendicular ($\theta _{LOS}\sim
90^{\circ}$). Knot A would be a head--on shock in this scenario. As we have
mentioned above, for this set of parameters additional acceleration has to
be provided to maintain the optical emission out to large distances from
the core. Meisenheimer et al. (1996) favor an unknown global acceleration
process to explain the constancy of the cutoff momentum. With these
assumptions, their model yields similar results to ours in that it
reproduces the small scale brightness variations on the basis of the
changes in cutoff frequency. Our model could thus be regarded as an
extension of their approach, putting it on the theoretical basis of
adiabatic expansions, with a different mechanism for providing the large
scale constancy of the spectrum.

\subsection{\em X-ray Emission}

An important result from the analysis above is that cooling longward of UV
energies is not important over the length of the jet --- the proper time is
reduced by a significant factor and the magnetic fields are small enough to
leave the spectral shape unchanged. This conclusion begs the question of
the origin of the X-ray emission detected by the {\em Einstein Observatory}
and {\em ROSAT}. Both observations show emission from the core/knot D
region and knot A.

\subsubsection{The Role of Particle Advection}
Ultra-high-energy particles, capable of radiating in the X-ray regime,
could be carried out from knot D, where X-ray emission is observed, to knot
A, and reaccelerated in the shock by the adiabatic compression mechanism
discussed above. For this to happen the spectrum would have to break
rather than cut off in the optical. The presence of a break instead of a cutoff
would not change our fits, as long as the break is located above the
frequency we fitted. We have calculated the behavior of particles with X-ray
emitting energies along the jet and found that for our best fit $B$-fields
cooling out to knot A will have produced a spectral cutoff at $\sim
10^{17}$ Hz -- which is where most of the {\em Einstein} HRI's sensitivity
lies. Since the $B$-fields we derived are upper limits, a lower field could
leave the distribution function unchanged even at such high
energies. Therefore this mechanism of producing X--rays is marginally
consistent with our model. It might also account for at least part of the
X--rays. Note that, as suggested by various authors (e.g.,
\cite{bir:book}), the X-ray emission could also be of non--synchrotron
origin altogether.

\subsubsection{Fermi Acceleration at Knot A}
The compression of a factor $\sim 3$ inferred from the analysis above
indicates that Fermi acceleration might be present in knot A, although the
moderate change in emissivity and the constancy in radio and optical
spectral index suggests that it might not be very efficient. It is possible
that Fermi acceleration occurs at parts of the shock only, resulting in a
particle distribution superposed of a compressed and a Fermi-accelerated
preshock distribution.

In order for Fermi acceleration to take place at all the shock has to
be subluminal, i.e., the intersection point of a given magnetic field line
and the shock front has to move with a speed smaller than the speed of
light, in which case we can find a frame in which the magnetic field is
perpendicular to the shock front. This is the case for fields not too closely aligned with the shock plane. In the nonrelativistic case the field
orientations leading to a superluminal shock are rare, and subluminal
shocks are the rule rather than the exception, so one would expect Fermi
acceleration to take place.

Because in relativistic shocks the percentage of superluminal field
orientations rises sharply with $\Gamma _{\rm shock}$, Fermi acceleration
should become less important. In this limit, most of the particle
acceleration would occur through the mechanism of ``shock drift
acceleration''. \cite{begelman} presented a theory of this process
valid in the relativistic case. They show that the adiabatic approximation
is still {accurate} in the limit of $\Gamma _{\rm p}\beta _{\rm p}\lesssim
1$, where $\Gamma _{\rm p}$ is the upstream Lorentz factor in the
perpendicular shock frame and $\beta _{\rm p}$ the corresponding
velocity. We have calculated this quantity for various obliquities and
field orientations appropriate for knot A and it seems to fall into the
desired range. For a disordered magnetic field we would have to average the
resulting spectrum over all possible field orientations. The more
superluminal the shock the less important would effects of Fermi
acceleration be. Relativistic corrections to shock drift {acceleration}
are only important for high $\Gamma _{\rm p}\beta _{\rm p}$, which, again,
depends on the field orientation. Averaging over all possible field
orientations would probably render these corrections unimportant.

Fermi acceleration both changes the shape of an incoming power law spectrum
and amplifies it. For an incoming electron spectrum of the form
$f(p)=A_{\rm 0}p^{-a}$ in the test particle limit (i.e., the pressure
provided by the accelerated particles is negligible), and a nonrelativistic
shock, the change of the spectrum depends on $s\equiv\frac{3r}{r-1}$, where
$r$ is the compression ratio. If $s<a$ the spectral index is changed to
$s$. If $s>a$ the slope remains unchanged but the spectrum is still
amplified by a factor $\frac{s}{s-a}$ (\cite{kirk}). The radio
spectral index is $\alpha\approx 0.5 - 0.65$, implying $a\approx 4 -
4.3$. To provide a boost in emissivity by a factor of $\approx 11$, $s$ has
to be $5.1 - 5.5$, implying a compression ratio of $\approx 2.2 - 2.5$
(assuming $\zeta=\frac{2}{3}$). Note that this is very close to the
compression ratio one derives for an adiabatic compression. The spectral
index produced by such a shock is $\alpha\approx 1 - 1.25$, consistent with
the observed $\alpha _{\rm optical}\approx1.2$. \cite{drury} showed that
the produced power law softens as one departs from the test particle
limit. Also, the simple treatment stated in this paragraph breaks down in
the case of relativistic shocks, where the spectral index is no longer a
simple function of the compression ratio.

Recent investigations by \cite{ballard} have shown that oblique
relativistic shocks can produce rather steep spectra, but other results
indicate that they might be more efficient in accelerating particles (i.e.,
producing flatter spectra) than their nonrelativistic counterparts
(\cite{kirkheavens}). \cite{sparks} find an optical--to--X-ray spectral
index of $\alpha _{\rm OX} \approx 1.4$ for the knot A region, which
seems to be consistent with low--efficiency acceleration.

If Fermi acceleration were present at the shock and effective enough to
change the spectral shape in the optical, it would no longer be feasible to
use the data from the whole jet to determine the magnetic field. Rather,
the same analysis could simply be carried out seperately for the pre-- and
post--knot A regions of the jet. We would then have to make an estimate of
the shock strength based on the known parameters in order to determine the
ratio of pre-- to post--shock Lorentz factor $\Gamma$. Even in this
case we would need Lorentz factors of order $\Gamma_{\rm jet} \approx 3$ to
solve the cooling problem.

However, based on the observed mild spectral changes, and the assumption
that pitch angle scattering is strong, we conclude that Fermi
acceleration, if present, will not be efficient enough to affect the
spectrum below the cutoff. {Inefficient Fermi acceleration might very
well be present in knot A, producing the X-ray emitting particles
observed. Prediction of the produced high--energy spectral index has to
wait for more conclusive results on Fermi acceleration at relativistic
oblique shocks.}

\subsection{\em Polarization}
\label{sec:polar}

An important complication to the treatment above is the fact that the
magnetic field will not be completely disorganized --- polarization
measurements show that in some regions a homogeneous component is
present. In fact, it is possible that cancellations between regions
with homogeneous fields but different orientations occur along the line of
sight (\cite{meis:polar}). In such a case, the assumption of disorganized
fields, leading to $\zeta \approx \frac{2}{3}$, is no longer
justified. However, since the impact that the parameter $\zeta$ has on the
fit is minor (Fig.\ \ref{fig:Berr}), we feel this caveat is not very severe
and merely mention this complication here.

In addition to the unknown orientation of the jet itself,
the field orientation is also unknown. Since synchrotron emission depends
on the magnetic field orientation to the line of sight $\vartheta$ as
$\sin^{1+\alpha}{\vartheta}$, it can be strongly peaked away from the field
direction. The cutoff frequency also depends on $\sin{\vartheta}$. Because
the cutoff frequency is determined by sampling all regions along the line
of sight, it is not obvious which value to choose for $\vartheta$.
Fortunately, in a domain of disordered field the regions with the field
oriented close to perpendicular will contribute most of the flux, thus the
error we make by setting $\sin{\vartheta}$ to 1 will not be too large. Note
that we used a value of $\vartheta=53^{\circ}$ in calculating the pressure,
the appropriate average of $\sin^{1.5}{\vartheta}$ over $4\pi$ sterad.

Knot A shows a polarization of order $35\%$ and a field orientation
close to the shock plane. Neglecting relativistic effects, the small amount
of upstream polarization, and the fact that shear might reduce the
compression of the field, we can use the approximate formula
(\cite{hughes})
\begin{equation}
\label{eq:polar}
\Pi\approx \frac{\alpha +1}{\alpha +
	\frac{5}{3}}\cdot\frac{(1-r^{-2})\cos^{2}{\epsilon}} 
	{2-(1-r^{-2})\cos^{2}{\epsilon}},
\end{equation}
where $\Pi$ is the fractional polarization, $r$ is the proper compression
ratio, and $\epsilon$ is the line of sight angle from the shock plane, to
obtain a lower limit on the compression ratio in knot A of $r\approx 2$,
valid for the inferred range of viewing angles with respect to the shock
plane (BB). This is consistent with the compression ratios inferred
above. Equation (\ref{eq:polar}) also provides an upper limit of
$35^{\circ}$ on $\epsilon$; however, the shock is assumed to be oblique, so
we cannot use this result to constrain $\theta_{LOS}$.

\subsection{\em Particle Acceleration Radius}

Having an estimate of both magnetic field and cutoff momentum at the
injection radius $r_{\rm 0}$, we can now try to determine where the actual
particle acceleration has to occur. We assume some radial dependence for
the magnetic field in the {\em inner} portion of the jet (i.e., smaller
than $0.5''$), for example a power law: $B\propto r^{-\sigma}$. Furthermore
we make the simplifying assumption of a constant $\Gamma_{\rm jet}$. By also
taking $B$ to be proportional to $\varrho^{\zeta}$, which determines the radial
dependence of $\varrho$, we can invert equation (\ref{eq:solution}) and solve
for the radius at which the cutoff momentum approaches infinity, in other
words, the minimum radius inside of which acceleration has to occur:
\begin{equation}
r_{\rm acc}=r_{\rm 0}\left(1+\frac{1}{A\cdot r_{\rm 0}}\right)^{-\lambda}
\label{eq:racc}
\end{equation}
where $A=\frac{4p_{\rm 0}B_{\rm 0}^{2}e^{4} \lambda}{9 \Gamma_{\rm jet} c^{7}
m_{\rm e}^{4}}$, and $\lambda\equiv\frac{1}
{(2+\frac{1}{3\zeta})\sigma-1}$. The acceleration radius $r_{\rm acc}$
approaches zero if $A\cdot r_{\rm 0}\rightarrow -1$. It is obvious that the
estimate of $r_{\rm acc}$ depends critically on the value of $\sigma$. In
Figure ~\ref{fig:racc} we have plotted $r_{\rm acc}$ as a function of
$\sigma$ for $\theta_{\rm LOS}=25^{\circ}$ and $\Gamma_{\rm jet}=3$. In the
same figure we have also plotted $r_{\rm acc}$ for the case in which B no
longer scales like $\varrho^{\zeta}$ --- in this case we have taken
$\varrho\propto r^{-2}$ and used the same values for $\theta_{\rm LOS}$ and
$\Gamma_{\rm jet}$ (dashed line).

If the jet expands at constant opening angle and with uniform $\Gamma_{\rm
jet}$ , the decline of B with radius should correspond to $\sigma \leq 2$,
since the density scales like $r^{-2}$ and dissipative effects will
probably limit the rate of decline. Adopting an upper limit on the magnetic
field at $0.01\ \rm pc$ of $B \leq 0.1\ \rm G$ (\cite{Reynolds}) limits
$\sigma$ to values smaller than $1$. For $\Gamma_{\rm jet}=3$, $\sigma=1$,
and $\zeta=\frac{2}{3}$, this constrains the acceleration radius to be
$r_{\rm acc}\leq 10\ \rm pc$, or $0''.06$. On the other hand, the radius of
acceleration cannot lie inside the Schwarzschild radius of the central
black hole, which is of order $10^{-4}\ \rm pc$ for a $10^{9}\ \rm
M_{\odot}$ black hole.

The dependence of $r_{\rm acc}$ on $\sigma$, $\zeta$, and the core
magnetic field is too strong to make any detailed predictions about where
the acceleration actually has to occur. However, the $r_{\rm acc}$--curve
is rather flat throughout most of the possible range for $\sigma$, which
suggests that the most plausible value for $r_{\rm acc}$ falls between $1$
and $10$ pc. This is intriguingly far away from the central engine.

\section{CONCLUSION} 
\label{sec-conclusion}

We have proposed that the apparent lack of synchrotron cooling in the M87
jet most likely indicates the presence of a sub-equipartition magnetic
field.  While the total (particle + magnetic) pressure needed to explain 
the observed synchrotron emissivity is uncomfortably high in the
nonrelativistic limit, Doppler beaming effects consistent with bulk Lorentz
factors in the range $2 - 5$ lower the pressure requirements considerably.
Fluctuations in the synchrotron emissivity and spectral cutoff frequency
are consistent with adiabatic changes in the magnetic field strength and
particle energies that accompany compressions and rarefactions along the
flow --- Fermi acceleration along the flow is not necessary to explain the
observations.

The knots are identified with relatively weak shocks, as inferred from
other data by \cite{bb}. The first-order Fermi acceleration expected to
occur at such shocks, if any, would generate a particle energy distribution
steeper than the $n(E) \propsim E^{-2}$ needed to produce the
radio--to--optical synchrotron spectrum. Thus, effects of particle
acceleration along the jet might be apparent only shortward of the cutoff
frequency, e.g., in the X-ray band.  However, a disordered magnetic field
or a ``superluminal'' field orientation with respect to the shock front
(Begelman \& Kirk 1990) could further reduce the efficiency of Fermi
acceleration, hence we should continue to regard the origin of the X-ray
emission as unknown.

The pressure estimates we derive are consistent with the assumption that
the M87 jet is embedded in a moderately overpressured bubble, as suggested
by \cite{bb}. As a result, it seems that the set of parameters we have
suggested above can solve the cooling problem of the M87 jet, as
sub-equipartition fields are able to explain both the behavior of the
cutoff frequency and the confinement of the jet.

Using derived values for the magnetic field and the cutoff momentum at
$r_{\rm 0}$ we can put an upper limit of $10\ \rm pc$ on the radius at
which most of the particle acceleration occurs. Explaining why the
acceleration is confined to a particular scale (which may be quite large
compared to the size of the central black hole) poses an interesting
problem for future work.

The method we have developed should be applicable to other radio sources as
well, as long as propagation effects on the particle momentum distribution
can be treated as adiabatic. Thus, as soon as comparable data becomes
available for other jets, it would be easy to apply the same analytic
techniques to them and obtain limits on the magnetic field and the pressure
in these sources. 

The analysis of the M87 jet we carried out in this paper could be improved
by better deprojection models of the jet. Higher quality spectral data
both in the near infrared and the X-ray would help to verify the assumption
of constancy of the underlying spectral shape and decrease the uncertainty
in the location of spectral features (the knowledge of which is essential
for this technique to work). {\em HST/NICMOS} will be ideally suited to
mapping out the jet properties in the near infrared, a spectral region in
which the run of the spectral index is still fairly poorly determined.
{\em AXAF}, providing imaging spectroscopy at sub-arcsecond resolution,
will be the ideal tool to obtain more detailed information about the X-ray
spectrum and emissivity along the jet, hopefully allowing us to determine
the radiation mechanism at high energies.

The model we have described above fits in nicely with the new picture of
M87 that has emerged in several recent papers. A moderately relativistic
jet and magnetic fields somewhat below equipartition can explain the lack
of cooling observed at optical wavelengths. Oblique shocks, weak enough to 
render Fermi acceleration unimportant, can explain both the small scale
variability of the cutoff frequency and the brightness variations via
adiabatic compressions. We believe this model can tie together most of the
observations available today and will take us closer to the true
nature of the jet.

\acknowledgments

We are grateful to Klaus Meisenheimer for an exceptionally thorough
and detailed referee report, which helped us to improve the paper. Klaus
Meisenheimer and Bill Sparks also provided us with the observational data
files that we used to fit our models. This work has been partially
supported by NSF grants AST-95 29175 and AST-91 20599 and a Fulbright
UP95/96/97 grant.

\clearpage

\figcaption[fit.eps]{Measured cutoff frequency along the jet
from \cite{meis96} (vertical lines with error bars). Solid line: best fit
curve as calculated from the emissivity measured by \cite{sparks} for
$\Gamma_{\rm jet}=1.1$, $\zeta =\frac{2}{3}$, and $\theta _{\rm
LOS}=90^{\circ}$; dashed line: best fit curve for the case of equipartition
calculated from \cite{sparks} data for the same set of
parameters. \label{fig:freqfit}}

\figcaption[contour.eps]{Contour plot of the $\chi ^{2}$ values as a
function of $B_{\rm mean}$ and the injection cutoff momentum  $P_{0}$
for $\theta =25^{\circ}$ and $\Gamma _{\rm jet}=3.0$. Shown are the
contours corresponding to a reduced $\chi^{2}$ of 45 and integer
multiples of this value. The shaded line on the right of the plot shows the
equipartition $B$-field for this set of parameters. \label{fig:contourplot}}

\figcaption[bfield.eps]{$B$-field averaged along the jet between $0''.5$
and $22''$ as a function of $\Gamma_{\rm jet}$ for $\theta_{\rm
LOS}=15^{\circ}$ (solid line), $\theta_{\rm LOS}=20^{\circ}$ (dotted line),
$\theta_{\rm LOS}=25^{\circ}$ (dashed line), and $\theta_{\rm
LOS}=30^{\circ}$ (dashed-dotted). The grey region indicates the
jet--to--counterjet brightness ratio limit of 150 (dashed grey boundary) - 380
(solid grey boundary, \cite{bir:ratio}).
\label{fig:Bmean}}

\figcaption[rfield.eps]{Ratio of the best--fit $B$-field to the
respective equipartition $B$-field $B _{\rm eq}$ averaged between
$0''.5\ {\rm and}\ 22''\ {\rm for}\ \zeta=\frac{2}{3}$ as a function of
$\Gamma_{\rm jet}$. The different values of $\theta _{LOS}$ are
labeled similarly to Fig. (\ref{fig:Bmean}). The grey region shows the
jet--to--counterjet brightness limit (dashed grey boundary: 150, solid grey
boundary: 380). \label{fig:Brels1p5}}

\figcaption[error.eps]{Fractional deviation $\frac{\Delta B}{B}$ of the
best fit $B$-field (averaged between $0''.5$ and $22''$ for {\em a.} two
different values of $\zeta$: $\left(\frac{\Delta B}{B}\right)_{\zeta}
\equiv{\frac{|B_{\zeta_{1}} - B_{\zeta_{2}}|}{B_{\zeta_{1}}}}$, where
$\zeta_{1}=\frac{2}{3}$ and either $\zeta_{2}=\frac{1}{15}$ (hatched, solid
boundary) or $\zeta_{2}=1$ (hatched, dotted boundary); {\em b.} a uniform
jet, compared to a jet slowing down at knot A: $\left(\frac{\Delta{B}}
{B}\right) _{\Gamma} \equiv\frac{|B_{\rm uniform} - B_{\rm break}|}{B_{\rm
uniform}}$, where $B_{\rm break}$ is the best fit average $B$-field for a
jet slowing down from $\Gamma_{jet,A-}$ to $\Gamma_{jet,A+}$ at knot A, and
$B_{\rm uniform}$ is the best fit average field for a uniform jet with 
$\Gamma_{jet} =\Gamma_{jet,A-}$ (dark grey region, short dashed boundary)
and $\Gamma_{jet}=\Gamma_{jet,A+}$ (light grey region, long dashed
boundary). The latter is plotted versus $\Gamma_{jet,A+}$; {\em c.} two
different 2cm radio spectral indices: $\left(\frac{\Delta
B}{B}\right)_{\alpha} \equiv {\frac{|B_{\alpha_{1}} - B_{\alpha_{2}}|}
{B_{\alpha_{1}}}}$ for our standard value $\alpha_{1}=0.5$ and
$\alpha_{2}=0.65$, shown as the black area. (The width of the band
corresponds to $\theta_{LOS}$ between $15^{\circ}$ and $30^{\circ}$ in each
case.)\label{fig:Berr}}

\figcaption[pressure.eps]{Total pressure $p_{\rm mean}$ averaged along
the jet between $0''.5\ {\rm and}\ 22''$  as a function of $\Gamma_{\rm
jet}$. Labels according to Fig. (\ref{fig:Bmean}). For comparison, the
dashed-triple-dotted line shows the equipartition pressure for a
non-relativistic jet seen edge on (i.e., $\theta_{\rm
LOS}=90^{\circ}$). The hatched area shows the estimated ISM pressure (White 
\& Sarazin 1988) in M87. \label{fig:logpressure}}

\figcaption[rpress.eps]{The ratio of the best--fit particle pressure to the
respective equipartition value averaged along the jet as a function of
$\Gamma_{\rm jet}$. Labels according to Fig. (\ref{fig:Bmean}).
\label{fig:relpressure}}

\figcaption[racc.eps]{The minimum acceleration radius $r_{\rm acc}$,
as a function of $\sigma$ ($B\propto r^{\sigma}$) for $\Gamma_{\rm
jet}=3$ and $\theta_{\rm LOS}=25^{\circ}$ in the case of $B\propto
\varrho^{\zeta}$ (solid line). The dashed line shows $r_{\rm acc}$ for the
case of $\varrho\propto r^{-2}$ and the same values of $\Gamma_{\rm jet}$
and $\theta_{\rm LOS}$. The hatched area indicates the limit set by 10
Schwarzschild radii for the $\sim 10^{9}\ {\rm M}_{\odot}$ central black
hole. The dashed-dotted line shows the (arbitrary) injection radius $r_{\rm
0}=0''.5$ or $80\ \rm pc$. \label{fig:racc}}

\clearpage
\plotone{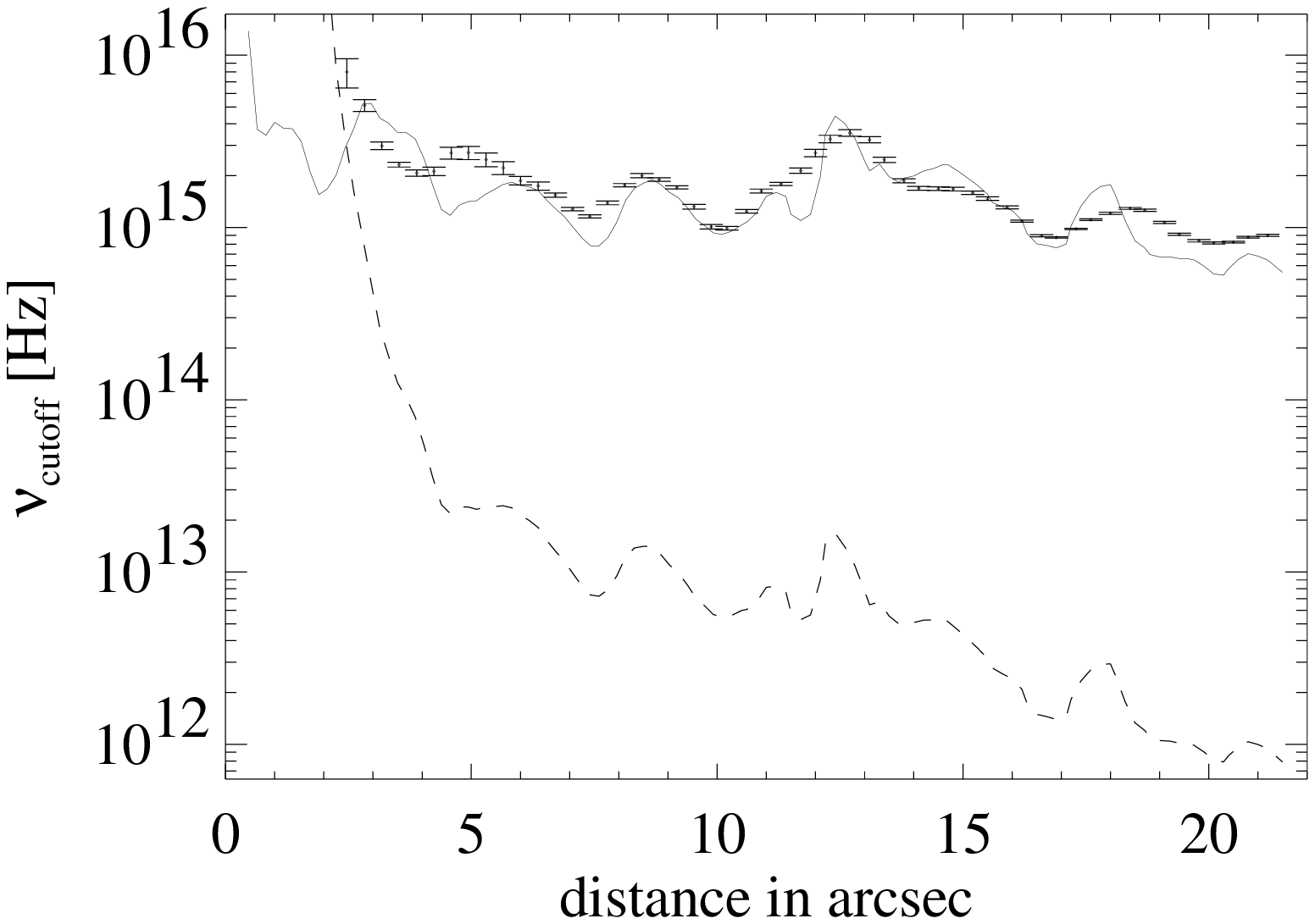}
\clearpage
\plotone{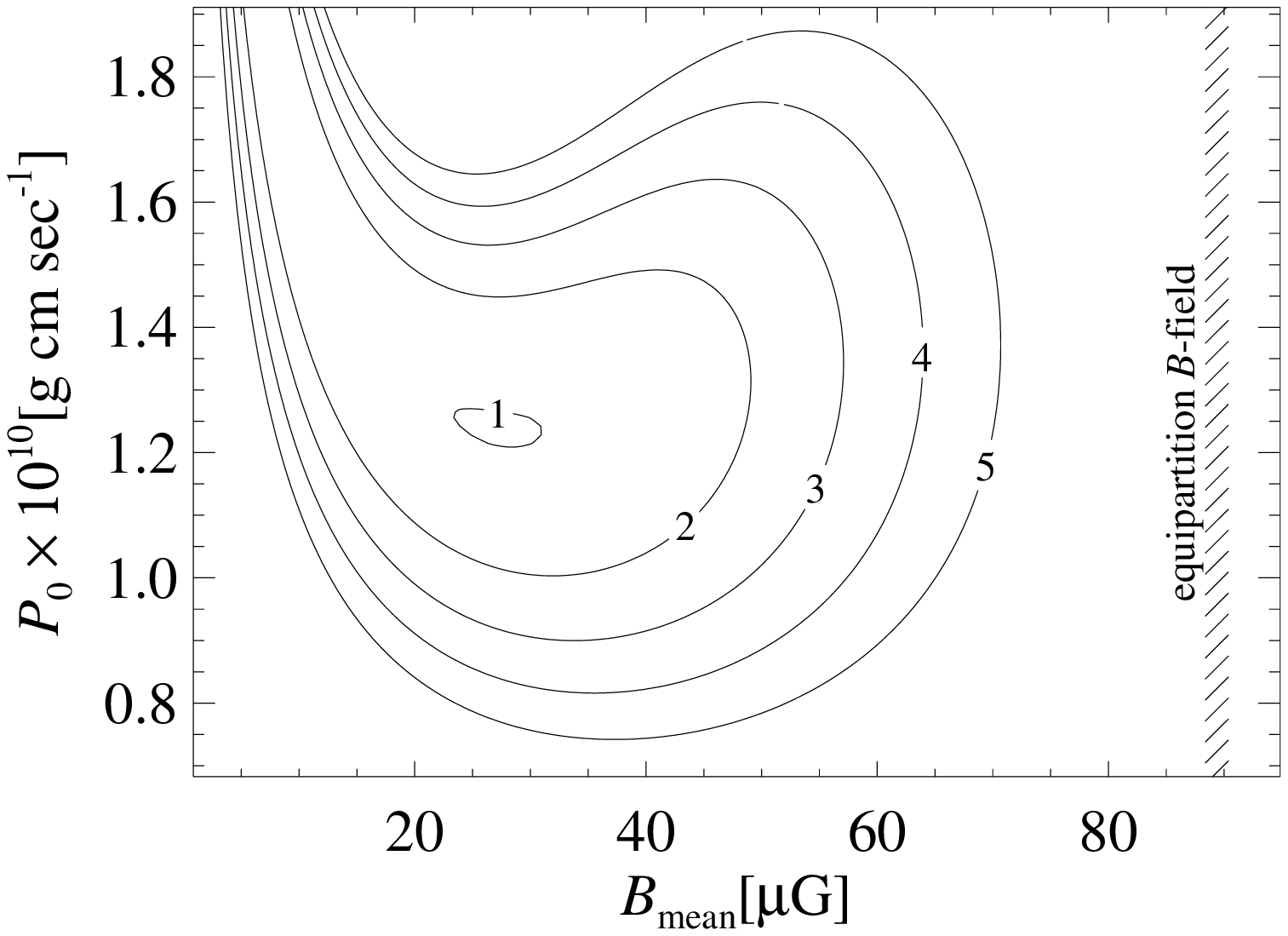}
\clearpage
\plotone{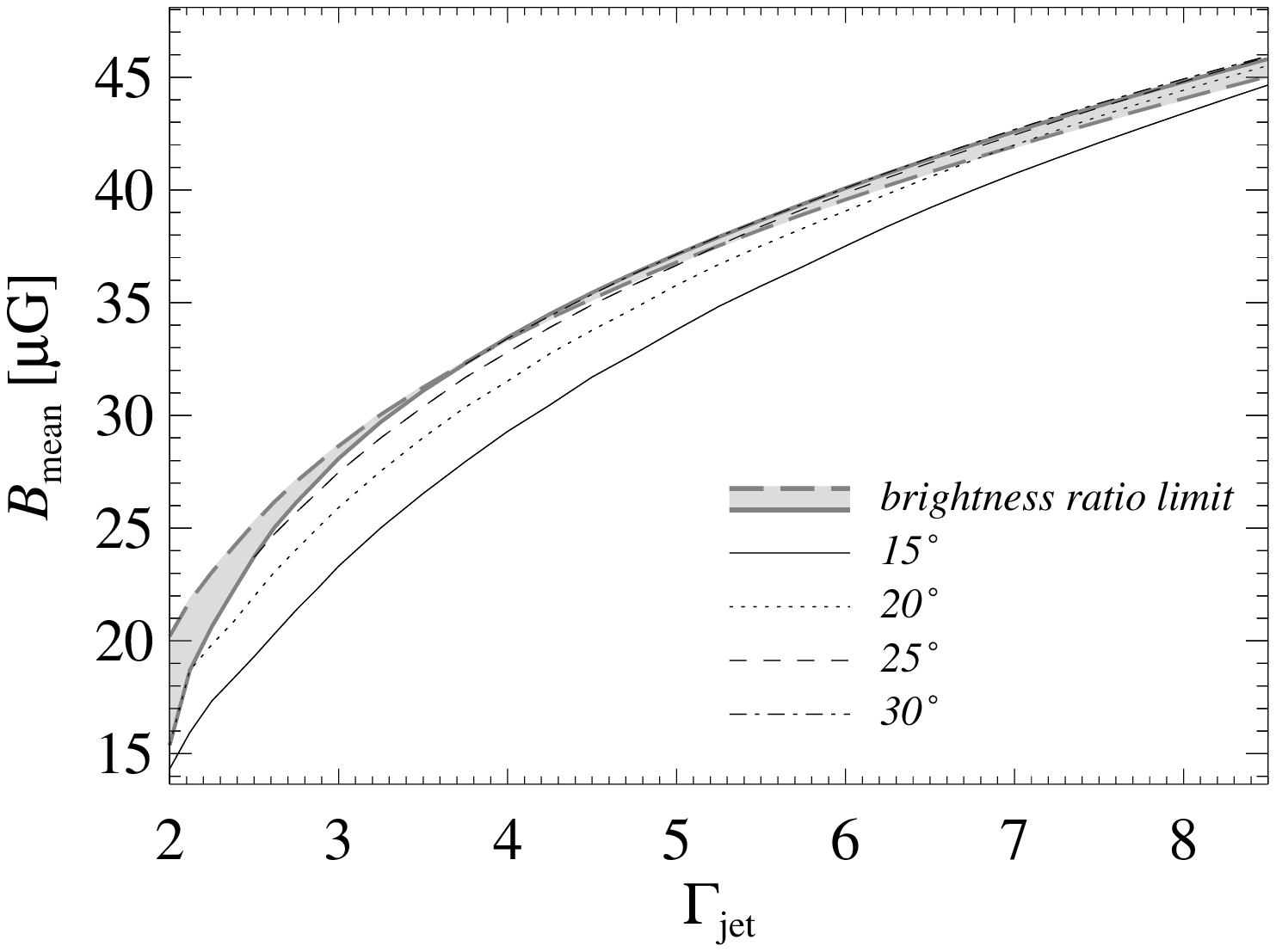}
\clearpage
\plotone{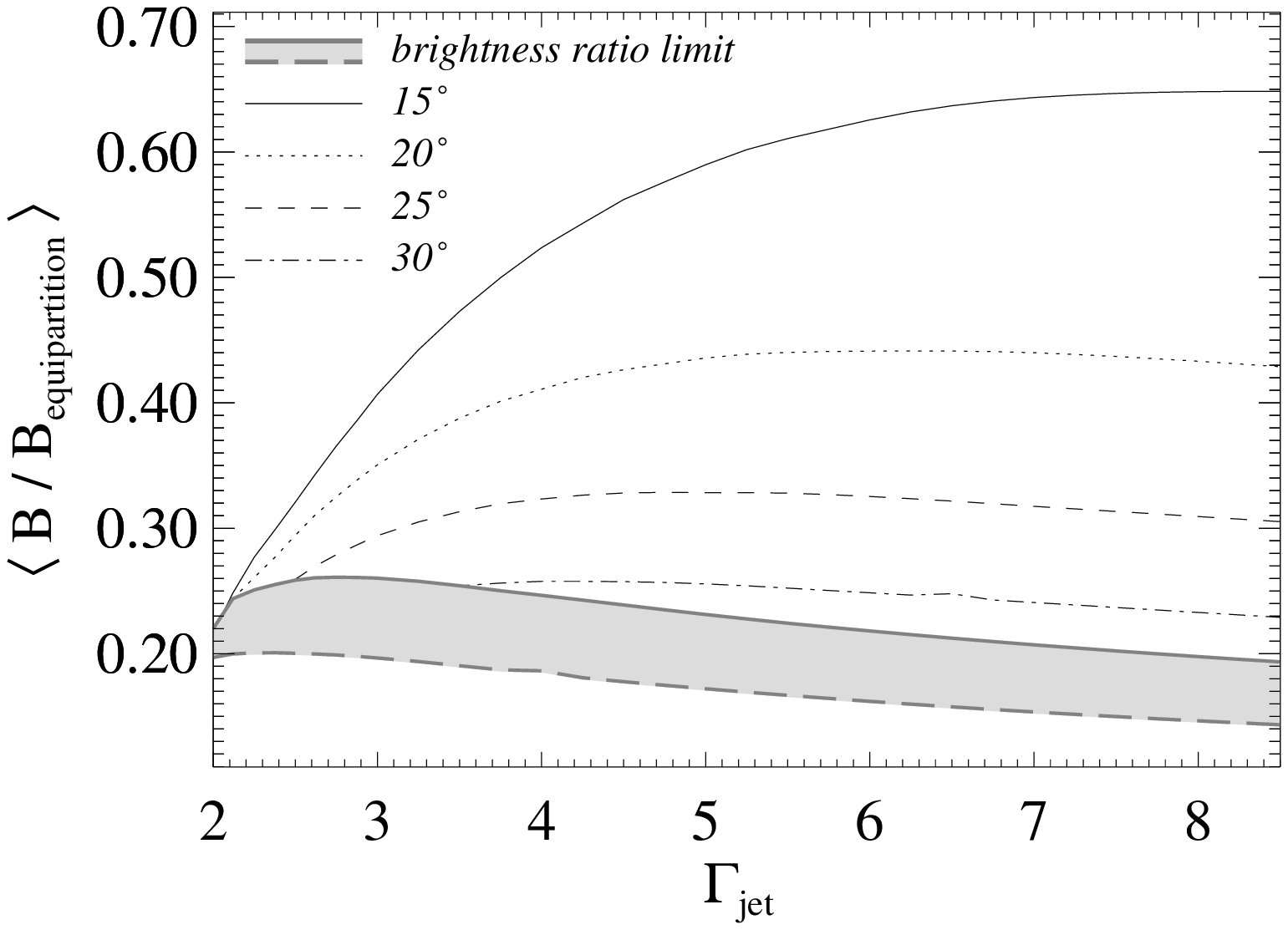}
\clearpage
\plotone{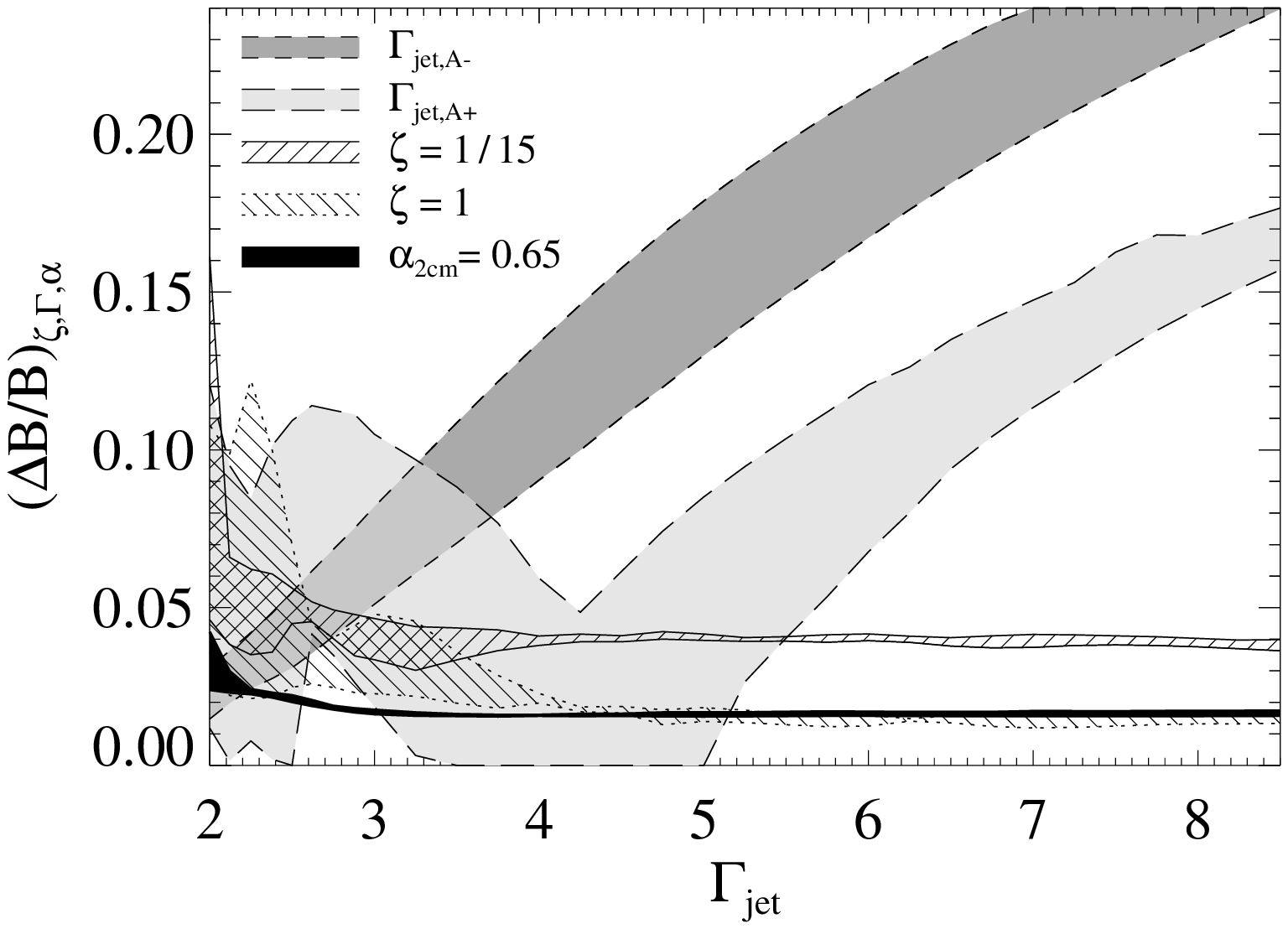}
\clearpage
\plotone{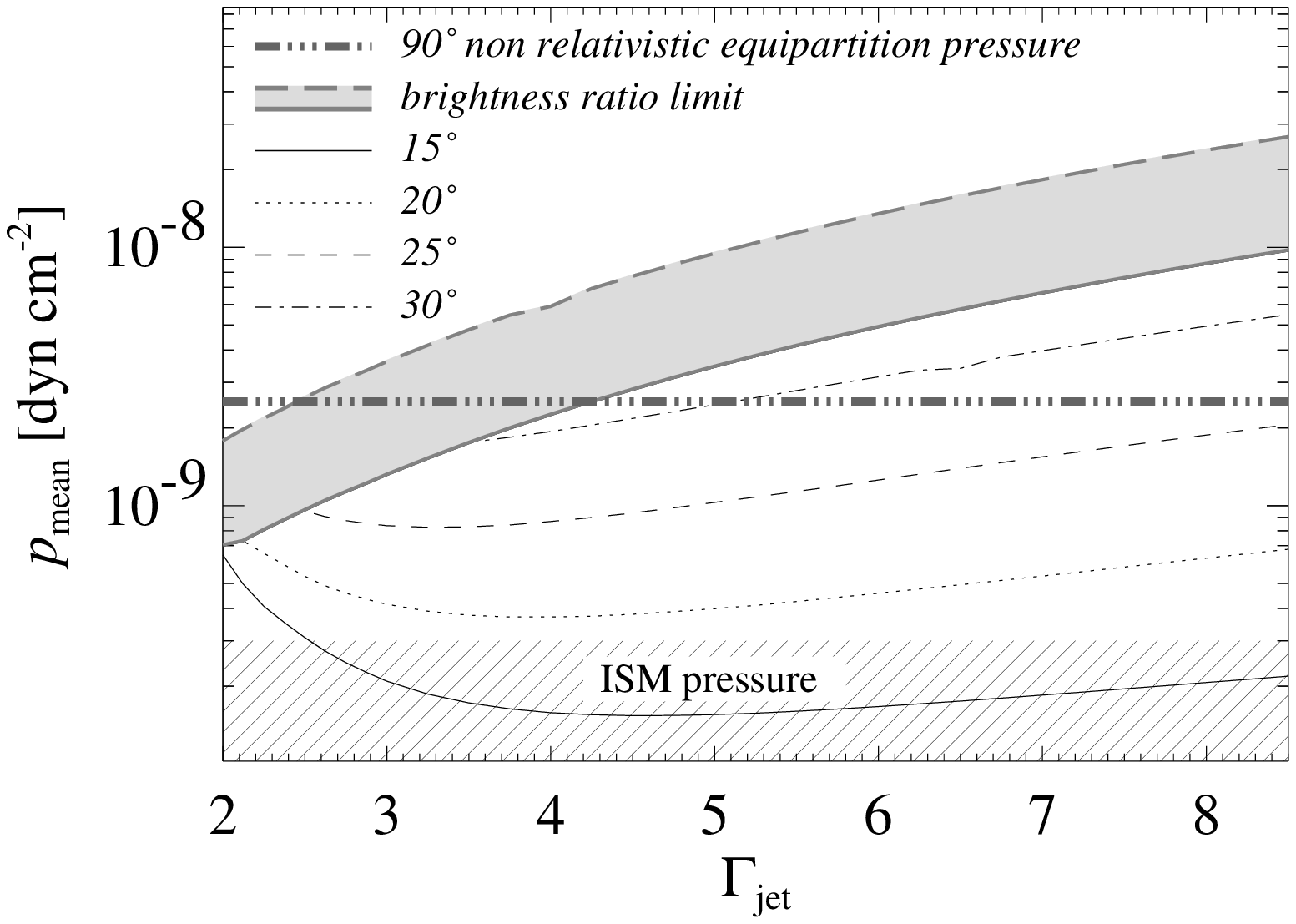}
\clearpage
\plotone{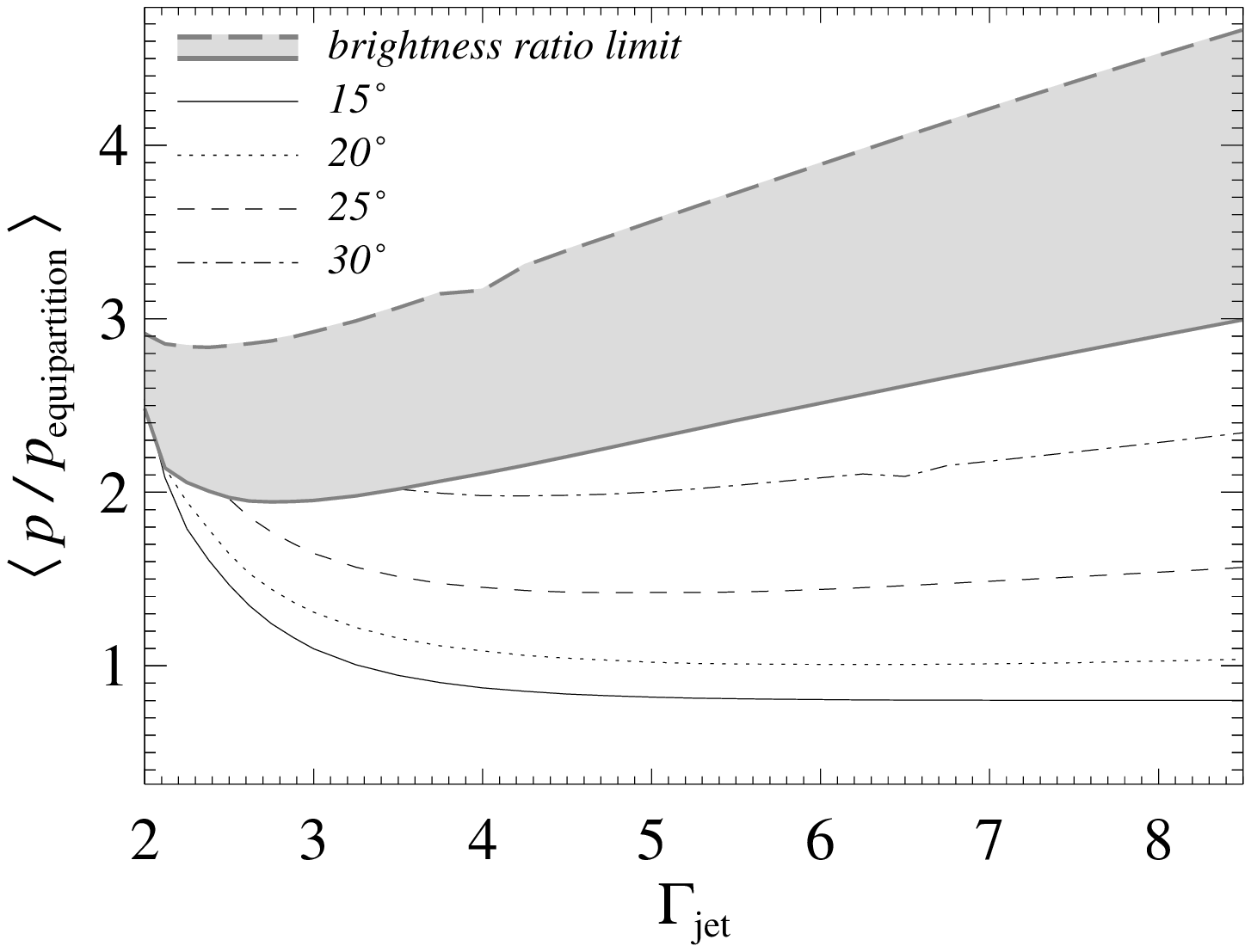}
\clearpage
\plotone{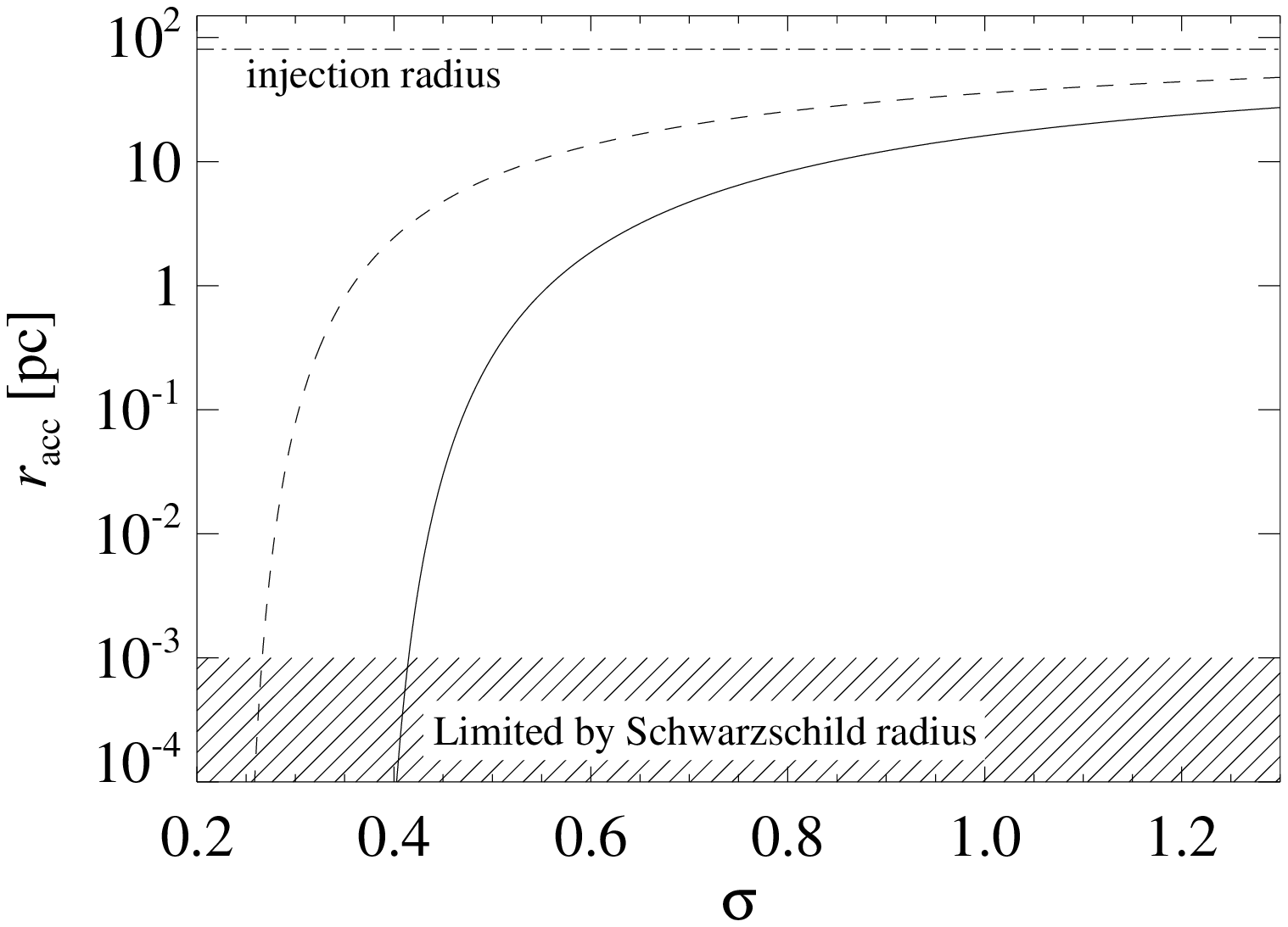}

\end{document}